\documentclass[3p,times,twocolumn]{elsarticle}

\usepackage{ecrc}

\volume{00}

\firstpage{1}

\journalname{Nuclear Physics B Proceedings Supplement}

\runauth{}

\jid{nppp}

\jnltitlelogo{Nuclear Physics B Proceedings Supplement}

\usepackage{amssymb}

\usepackage[figuresright]{rotating}

\begin{document}

\begin{frontmatter}



\dochead{}

\title{Recent developments in heavy flavor probes in lattice QCD}


\author{Anthony Francis}
\ead{afranc@yorku.ca}
\address{Department of Physics and Astronomy, York University, M3J1P3, Toronto, ON, Canada}

\begin{abstract}

The analysis of heavy flavor lattice correlation functions to obtain insights into transport phenomena and bound state dissociation patterns is a difficult and interesting challenge from the point of view of lattice QCD spectroscopy. In this contribution to "Hard Probes 2015", the recent developments and advances from different lattice efforts to determine the relevant spectral properties will be reviewed and discussed. The difficulties underlying this line of research will be highlighted and the obtained results assessed. Among others, results on the dissociation of charm systems and bottomonia, as well as the diffusion of heavy quarks will be addressed. 
\end{abstract}


\end{frontmatter}



\section{Hard probes in lattice QCD}

Heavy-ion collision experiments probe nuclear matter under extreme conditions. A detailed theoretical knowledge of the non-perturbative phenomena that underlie the processes governing the formation and evolution of these systems is therefore of fundamental importance.
Lattice QCD excels at accessing static, Euclidean-time quantities such as the equation of state \cite{Bazavov:2014pvz,Borsanyi:2013bia} and quark number susceptibilities \cite{Bellwied:2013cta,Bazavov:2013dta}. The real-time information on the other hand is hidden in the numerical lattice data and encoded in spectral functions $\rho(\omega,\vec p,T)$, which are difficult to extract. Among the quantities that are accessible only in this way are the complex heavy quark potential, heavy quark dissociation patterns and heavy quark diffusion coefficients. The recent advances on these quantities are reviewed in the following.

\section{Complex heavy quark potential}
\label{}

\begin{figure}[t!]
\centering
\includegraphics[width=0.45\textwidth]{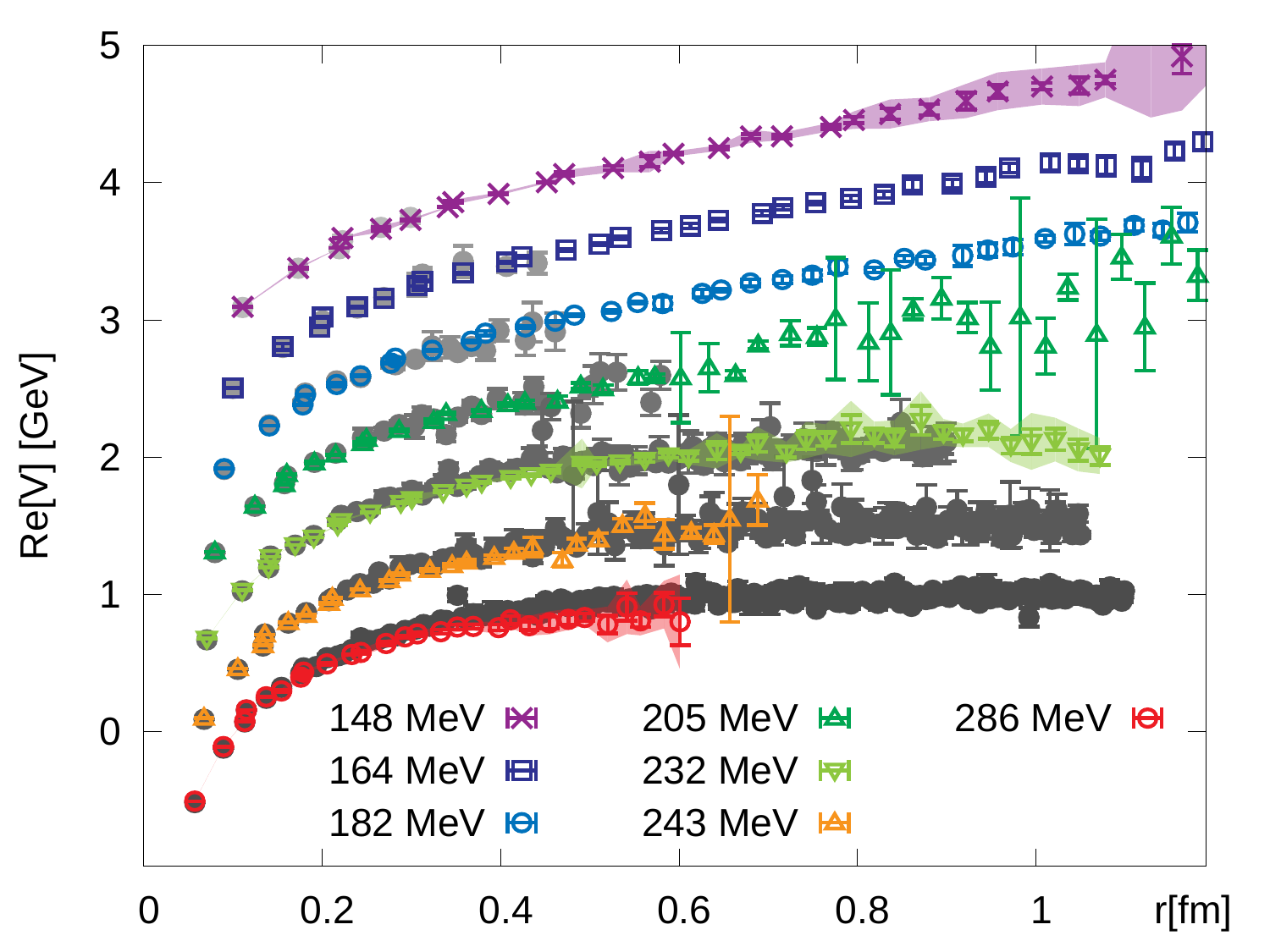}
\includegraphics[width=0.45\textwidth]{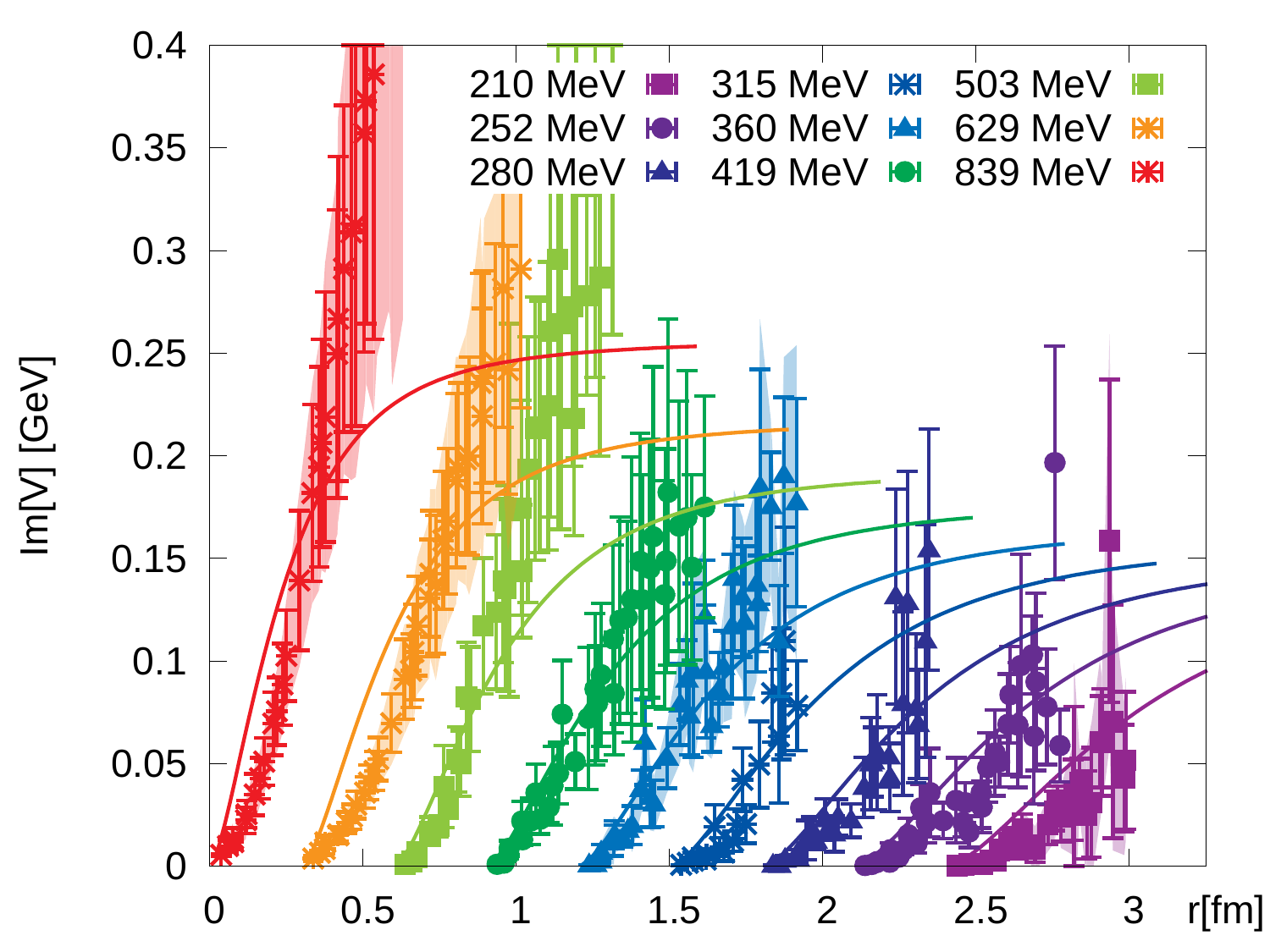}
\caption{The dynamical real part and quenched imaginary part of the complex heavy quark potential \cite{Burnier:2014ssa}. The grey points indicate results obtained from computing the free energies.}
\label{fig:hqpotential}
\end{figure}

In the past, the question of quarkonium dissociation was addressed by computing the heavy quark free energies on the lattice and combining the results with knowledge from potential models \cite{Kaczmarek:2005ui}. These calculations estimated heavy quarkonium dissociation temperatures up to $\sim 1.5T_c$. It is however not clear if the heavy quark free energy accurately describes the heavy quark potential, since the latter is a complex quantity at finite temperature \cite{Brambilla:2004jw,Laine:2007qy,Beraudo:2007ky,Brambilla:2008cx}. In \cite{Rothkopf:2011db} the authors laid out a strategy to compute the complex heavy quark potential from spectral functions of the thermal Euclidean Wilson loop:
\begin{eqnarray}
V(r)&=&\lim_{t\rightarrow\infty}\frac{i\partial_t W(t,r)}{W(t,r)} \nonumber\\
&=&\lim_{t\rightarrow\infty} \frac{\int d\omega\omega e^{-i\omega t}\rho(\omega,r) }{\int d\omega e^{-i\omega t}\rho(\omega,r) }~~,
\end{eqnarray}
thereby enabling an ab initio calculation using lattice methods. This computation was carried out in \cite{Burnier:2014ssa} on anisotropic, quenched and isotropic, dynamical $N_f=2+1$ staggered (asqtad) ensembles. An analytic Ansatz to fit the acquired data based on a single temperature dependent parameter, the Debye mass, was later added in \cite{Burnier:2015nsa}. The results are shown in Fig.~\ref{fig:hqpotential} for the real part in the dynamical setup (top), where the grey points indicate the results from computing the free energies, and the imaginary part in quenched theory (bottom). Unfortunately in the dynamical case the signal for the imaginary part does not permit an estimate at this time. In the quenched setup the Debye masses extracted in \cite{Burnier:2015nsa} lie between $m_D\simeq850$MeV and $330$MeV in the Temperature range between $T\simeq840$MeV and $210$MeV. These values are consistently lower than those found from extracting the Debye mass from the free energies \cite{Maezawa:2007fc}. An update of the dissociation temperatures based on these results is not performed yet.

\section{Current-current spectral functions from lQCD}

To determine the dissociation temperatures of quarkonia, a more direct way is to reconstruct the spectral functions $\rho(\omega,\vec p,T)$ from mixed representation heavy current-current correlation functions calculated on the lattice:
\begin{eqnarray}
G(\tau,\vec p, T)&=&\int_0^\infty \frac{d\omega}{2\pi}\rho(\omega,\vec p,T) K(\omega,\tau,T) \label{eq:spf}\\
K(\omega,\tau,T)&=&\frac{\cosh(\omega(\tau -\beta/2))}{\sinh(\omega\beta/2)}\label{eq:kernel}~~,
\end{eqnarray}
where $\beta=1/T$. Once the spectral functions are reconstructed the dissociation temperature of quarkonia can be read off from the disappearance of bound state peaks with increasing temperature. The transport coefficients can be determined via Kubo formulas of the type:
\begin{equation}
	D= c^*  \lim_{\omega\rightarrow 0}\frac{\rho_{ii}(\omega,\vec p=0,T)}{\omega T}~~,
\end{equation}
where the proportionality constant $c^*$ depends on the system at hand, see e.g. \cite{Petreczky:2005nh,Meyer:2011gj}. 
The required lattice correlation function, in the local-local case, is determined by calculating:
	\begin{eqnarray}
	G_{\mu\nu}(\tau,\vec p)&=&\sum_{\vec x} G_{\mu\nu}(\tau,\vec x,T)\,e^{i\vec p\vec x}\\
	G_{\mu\nu}(\tau,\vec x)&=&\Big\langle \Big(\bar\psi \Gamma_\mu \psi\Big)(\tau,\vec x) \Big(\bar\psi \Gamma_\nu \psi\Big)^\dagger(0,\vec 0)\Big\rangle~~. \label{eq:corr}
	\end{eqnarray}	
Performing the necessary Wick contractions in Eq.~(\ref{eq:corr}) generally two terms emerge, a connected and a disconnected part. However, due to numerical cost the disconnected term is neglected in all the lattice calculations presented here. For heavy mesons the signal is generally strong enough that an additional smearing does not have to be applied, thereby simplifying the interpretation of results. This is dependent on the observable and is e.g. not the case fore baryons at finite temperature \cite{Aarts:2015mma}. Note, due to the parity mixing in correlators computed using the (relativistic) staggered action, spectral functions are usually reconstructed using Wilson-Clover type lattice actions, if not stated otherwise this will be tacitly assumed in the following.\\
Reconstructing the spectral function via computing the inverse transformation of Eq.~(\ref{eq:spf}) is a numerically ill-posed problem. To some extent any result therefore depends on the assumptions and systematics of the reconstruction technique. Additionally, the difficulty of estimating the systematic uncertainty of reconstructed spectral functions is increased due to the suppressive behavior of the Kernel in Eq.~(\ref{eq:kernel}).
Two approaches have been generally adopted to deal with the reconstruction problem, see \cite{Meyer:2011gj} for a review, and a third one has been recently introduced \cite{Brandt:2015sxa}. The first is the maximum entropy method (MEM) that invokes Bayes' theorem to determine the most probable spectral function given the data with its errors and a so called default model. Here, the systematic uncertainties are rooted in the dependence on the default model and also the underlying algorithm \cite{Burnier:2013nla}. In the second approach different phenomenologically motivated ans\"atze for the spectral functions are inserted into Eq.~(\ref{eq:spf}) and directly fit to the lattice data. The dominant systematic effects are therefore in the choice of the ans\"atze. A third option, the Backus-Gilbert method, aims at determining the spectral function as local as possible in the frequency $\omega$ given a certain resolution function. The finite width of the resolution function then constitutes the dominant source of uncertainty.

\section{Quarkonium spectral functions}

\subsection{Charmonium}

In quenched QCD, charmonium dissociation and the charm diffusion constant have been studied in \cite{Ding:2012sp}. The correlators were computed on large, isotropic quenched ensembles and the spectral function reconstruction was carried out using MEM with a rescaled kernel $K'(\omega,\tau,T)=\tanh(\omega\beta/2)K(\omega,\tau,T)$ and $\rho'(\omega,\vec p,T)=\rho(\omega,\vec p,T)/\tanh(\omega\beta/2)$, which cures the Kernel divergence as $\omega\rightarrow 0$ \cite{Aarts:2007wj,Engels:2009tv}.

\begin{figure}[t!]
\centering
\includegraphics[width=0.4\textwidth]{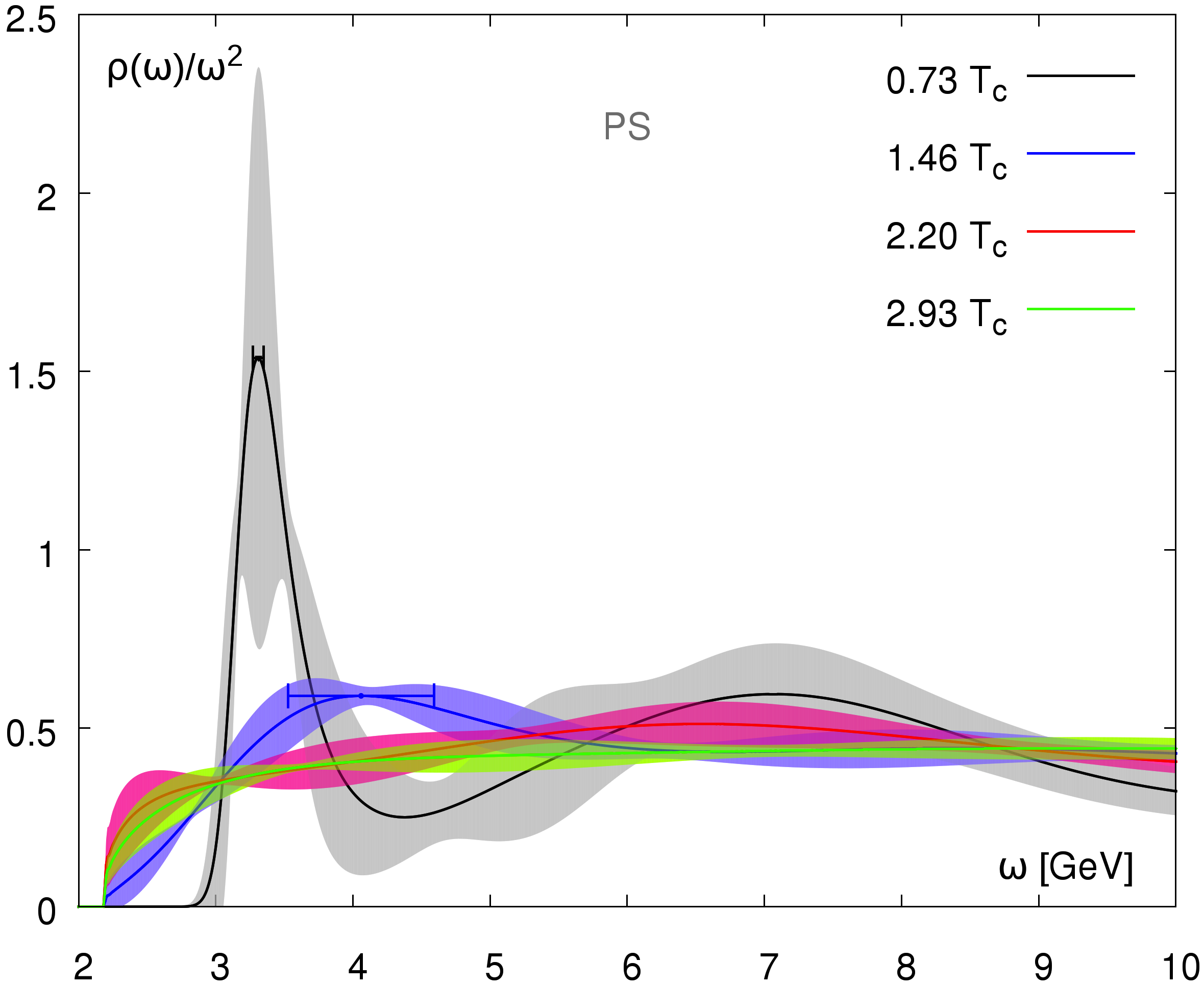}
\includegraphics[width=0.4\textwidth]{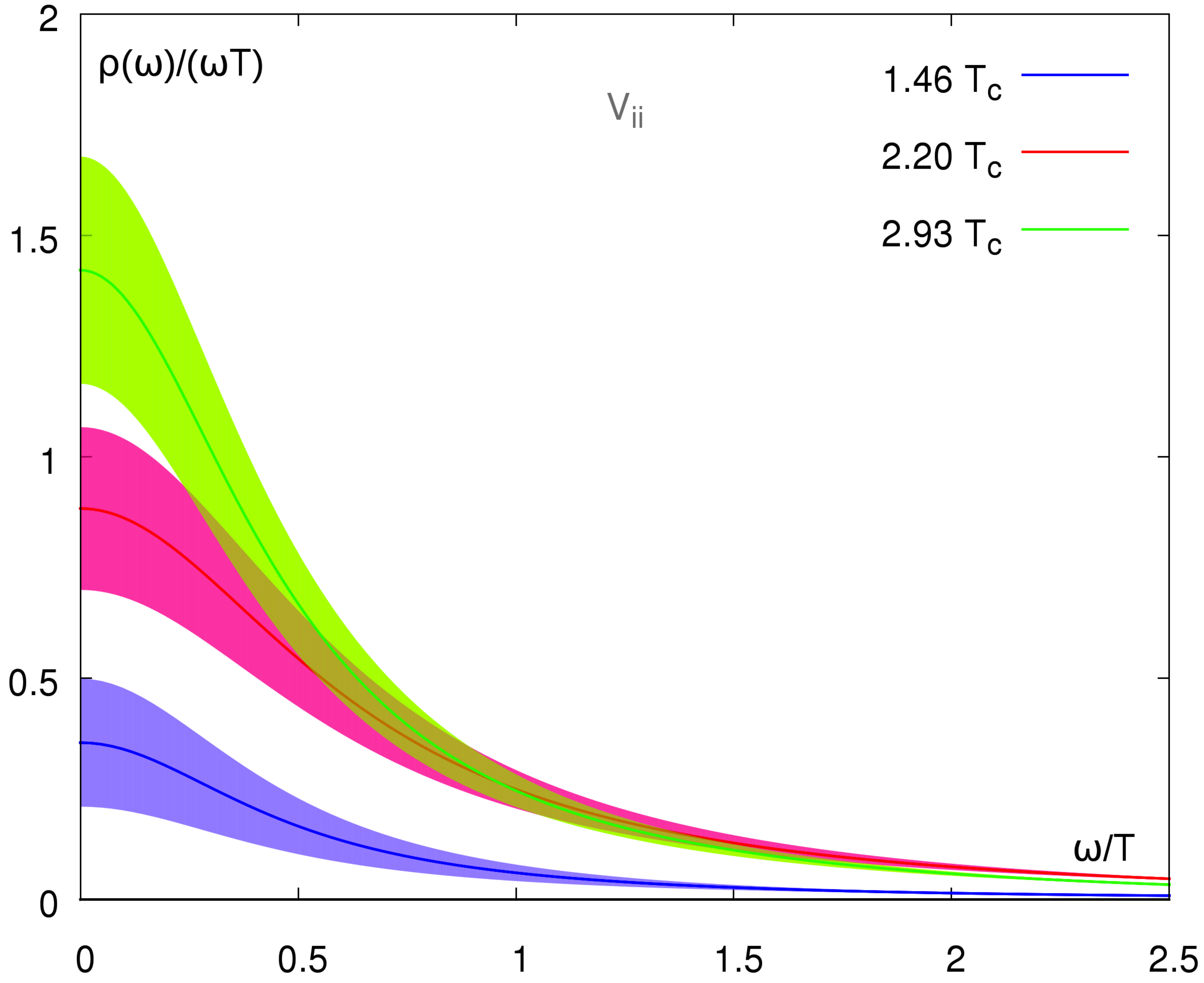}
\caption{Top: Charmonium spectral function $\rho(\omega,\vec p=0,T)/\omega^2$ in the pseudo scalar channel. Bottom: The low frequency region and transport peak $\rho(\omega,\vec p=0,T)/\omega T$ in the vector case. Both results from \cite{Ding:2012sp}.}
\label{fig:charmquench}
\end{figure}

\begin{figure}[h!]
\centering
\includegraphics[width=0.45\textwidth]{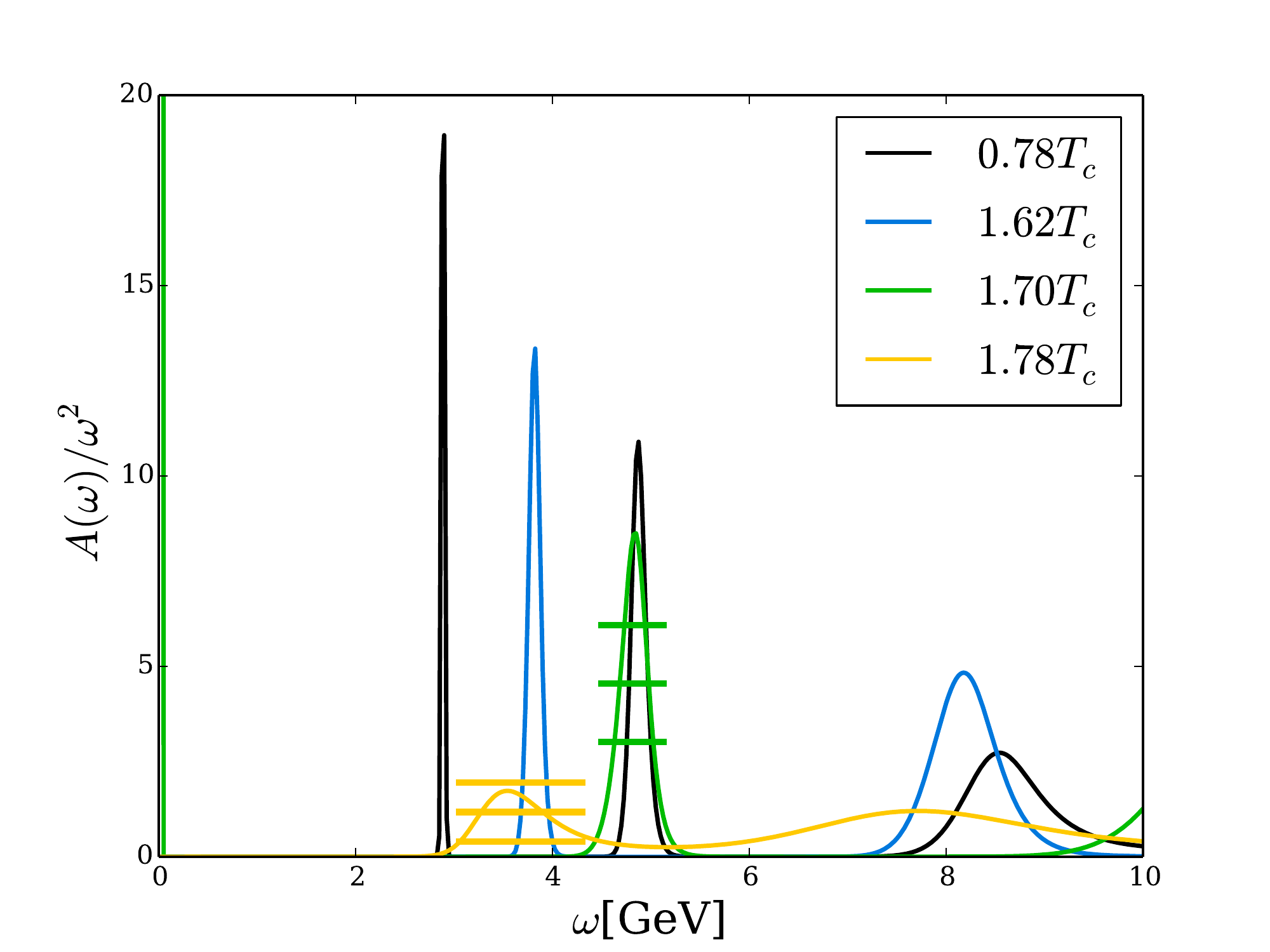}
\includegraphics[width=0.45\textwidth]{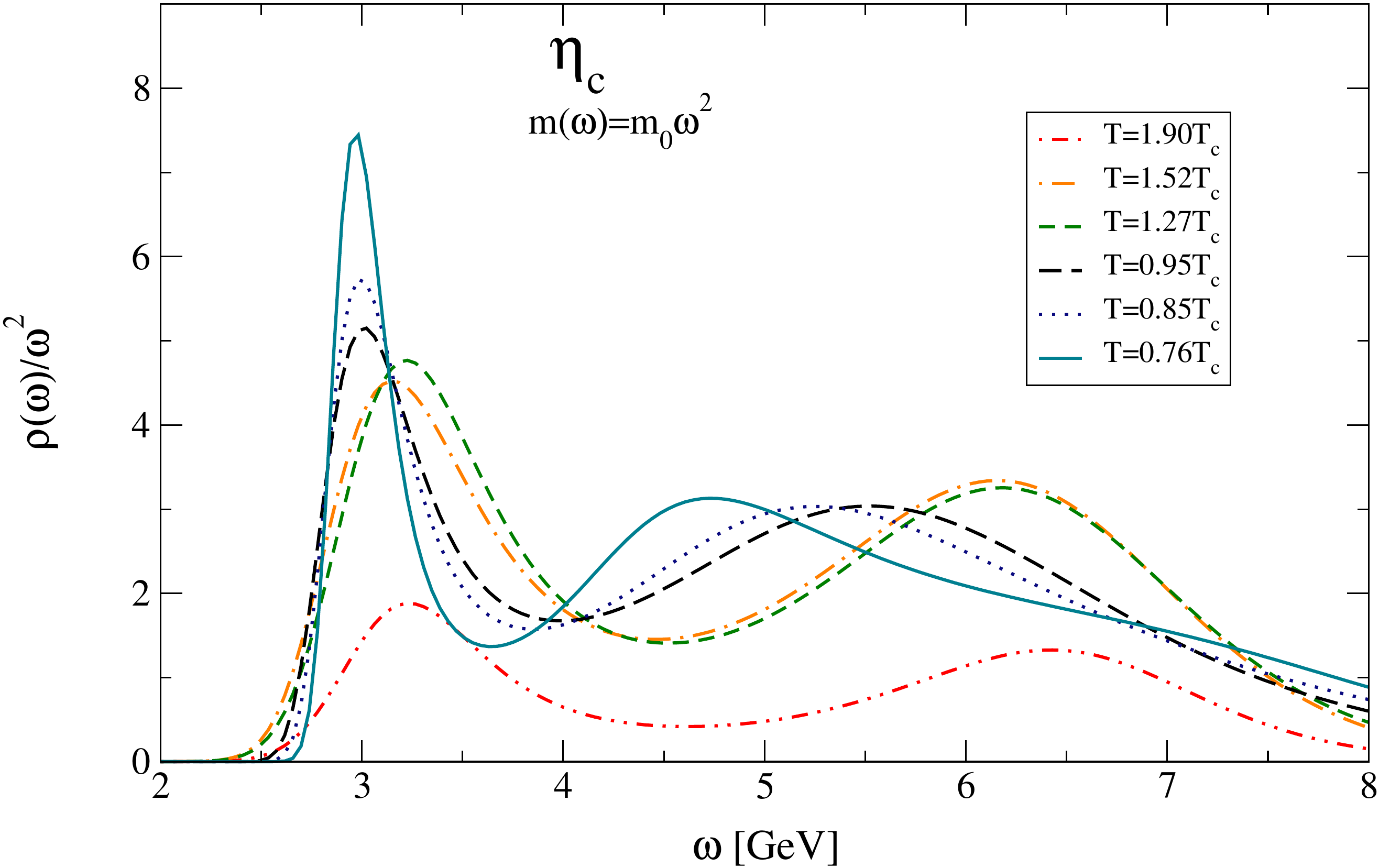}
\caption{Charmonium spectral functions $\rho(\omega,\vec p=0,T)/\omega^2$ in the pseudo scalar channel from recent quenched \cite{Ikeda:2014vca}(top) and dynamical \cite{Skullerud:2014sla}(bottom) studies.}
\label{fig:charmdyn}
\end{figure}

Representative results for the $\eta_c$ (pseudo scalar) spectral function are shown in Fig.~\ref{fig:charmquench} scaled as $\rho(\omega,\vec p=0,T)/\omega^2$ (top) and for the $J/\Psi$ (vector) case as $\rho(\omega,\vec p=0,T)/\omega T$ (bottom). The former highlights the dissociation of the bound state peak, while the latter highlights the transport region. In summary the diffusion coefficient was found to be $2\pi TD\sim 1...3$, while both the $J/\Psi$ and the $\eta_c$ peaks visible at $0.73T_c$ are suppressed in the spectral function around $1.5T_c$.
Away from quenched theory results on anisotropic lattices in $N_f=2$ theory at $m_{\pi}\simeq 500$MeV and MEM reconstruction \cite{Aarts:2007pk} suggested dissociation temperatures for both these particles around $2T_c$. Recent updates of these results in quenched \cite{Ikeda:2014vca} and dynamical \cite{Skullerud:2014sla} QCD have focused on the pseudo scalar, $\eta_c$, channel. This channel is more accessible, since a transport contribution is not expected and the dissociation should be easier to detect. The first study, \cite{Ikeda:2014vca}, 
uses anisotropic, quenched ensembles and an extended MEM to pin down the spectral function.
The second, \cite{Skullerud:2014sla}, uses $N_f=2+1$ anisotropic lattice ensembles with $m_{\pi}\simeq 400$MeV and also MEM for spectral function reconstruction.
Representative results of both of these studies are given in Fig.~\ref{fig:charmdyn} as $\rho(\omega,\vec p=0,T)/\omega^2$. In both cases a flattening of the spectral function is observed below $1.90T_c$, however there are no clear signs of dissociation in this temperature region.

\subsection{Bottomonium}

\begin{figure}[b!]
\centering
\includegraphics[width=0.45\textwidth]{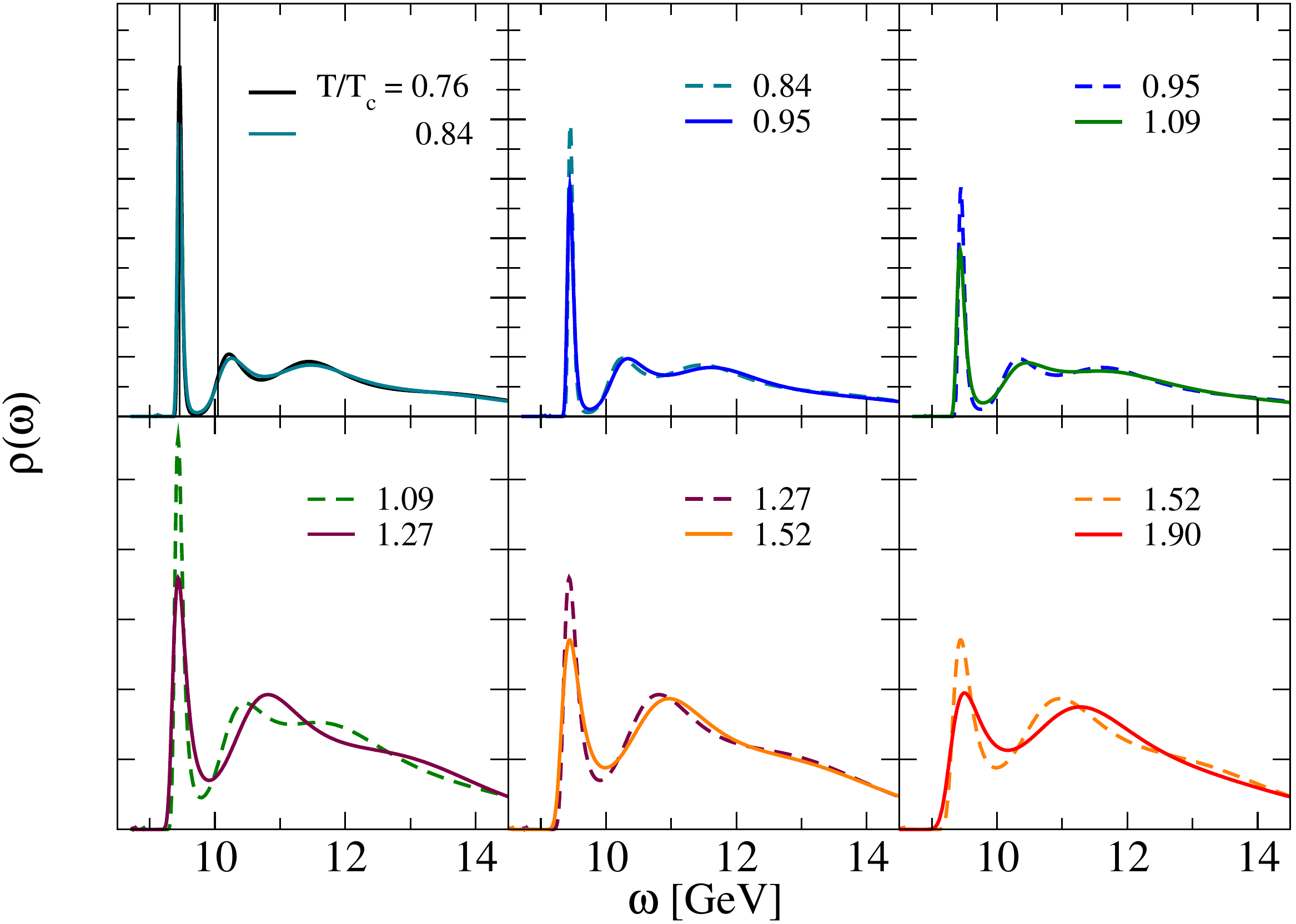}
\includegraphics[width=0.47\textwidth]{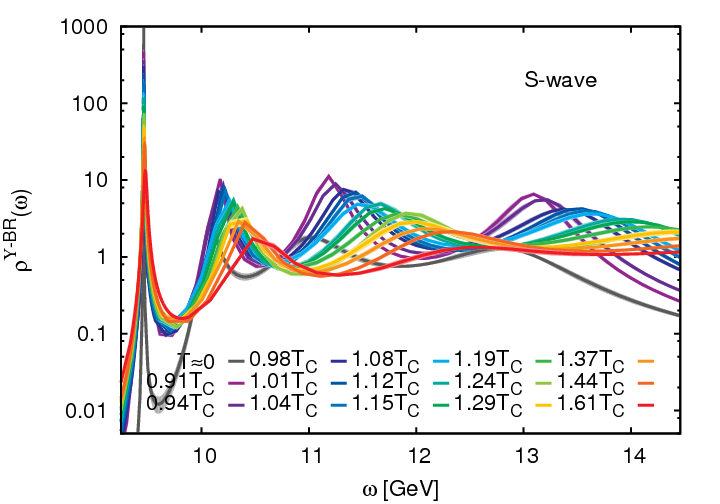}
\includegraphics[width=0.45\textwidth]{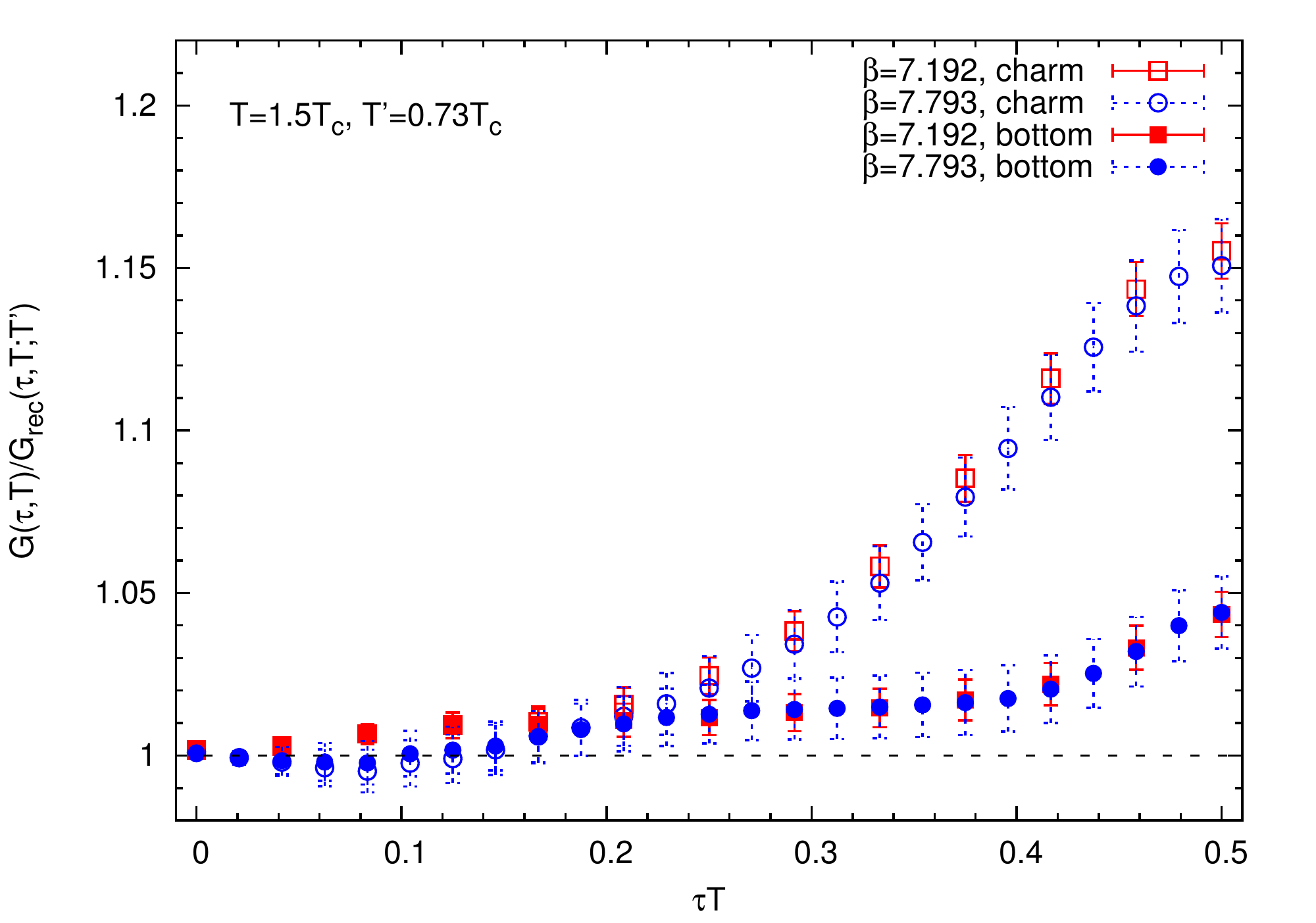}
\caption{Top: Bottomonium $\Upsilon$ spectral functions determined from NRQCD correlators on anisotropic $N_f=2+1$ lattice ensembles at $m_{\pi}\simeq 400$MeV via MEM reconstruction \cite{Aarts:2011sm}. Middle: The NRQCD bottomonium $\Upsilon$ spectral functions from isotropic, staggered ensembles at $m_{\pi}\simeq 160$MeV using an extended Bayesian reconstruction technique \cite{Kim:2014iga}. Bottom: Ratios of the thermal and the reconstructed correlators for charmonium and bottomonium in the vector channel in the relativistic approach on large quenched ensembles \cite{Ohno:2014uga}. }
\label{fig:bottomdyn}
\end{figure}

Accessing bottomonia from lattice QCD is hampered by the very small lattice spacings required to resolve the bottom particles without significant lattice effects. Although state-of-the-art quenched calculations have reached values of the cut-off that enable a fully relativistic approach to bottomonia \cite{Ohno:2014uga}, in the dynamical set up this is beyond reach. However, for bottom quarks the heavy quark scale may be integrated out to form a non-relativistic effective field theory, NRQCD. Here, the spatial lattice spacing acts as a short distance cut-off thereby requiring $m_b\,a_s\gtrsim 1$. Consequently it is a favourable approach for anisotropic and coarse lattice ensembles. Studies using NRQCD additionally have the advantage, that the transport region has been integrated out with the heavy quark mass scale. As in the pseudo scalar charmonium case the dissociation of bottomonia should therefore be more clearly visible in the lattice data. Representative results from a recent update \cite{Aarts:2014cda} of \cite{Aarts:2011sm} have been collected in Fig.~\ref{fig:bottomdyn}(top), in both cases the lattice ensembles are $N_f=2+1$ anisotropic with $m_{\pi}\simeq 400$MeV and MEM was used for spectral function reconstruction. The clear bound state peak in all ensembles suggests no dissociation of the $\Upsilon$ below $1.90T_c$. 
In a study using isotropic staggered (asqtad) ensembles and an extended Bayesian reconstruction technique \cite{Kim:2014iga}, these findings were confirmed up to temperatures below $1.61T_c$, see Fig.~\ref{fig:bottomdyn}(middle).\\
Unlike the NRQCD studies the aforementioned quenched study \cite{Ohno:2014uga}, has the advantage of being able to treat the charm and bottom systems on equal footing, while at the same time being able to access the transport region.
So far a spectral function reconstruction has not been published yet, however the associated ratios of the finite temperature over the reconstructed vector correlators are available and shown in Fig.~\ref{fig:bottomdyn}(bottom). The reconstructed correlator, i.e. the correlator that would be obtained at finite temperature, if the spectral function remained the vacuum one, can be computed without spectral function reconstruction by exploiting \cite{Ding:2012sp,Meyer:2010ii}:
\begin{equation}
G_{\rm rec}(\tau,T;0) = \sum_{m\in\mathbb{ Z}} G(|\tau+m\beta|,T=0)~~.
\end{equation}
For charmonium a large deviation from unity is observed for large distances, compatible with \cite{Ding:2012sp}. In the bottomonium case the ratio stays close to unity for the entire Euclidean time distance, indicating only little change in the bound state peak and the emergence of a transport peak, however without spectral function reconstruction firm conclusions are not yet possible.

\section{Heavy quark diffusion}

\begin{figure}[h!]
\centering
\includegraphics[width=0.45\textwidth]{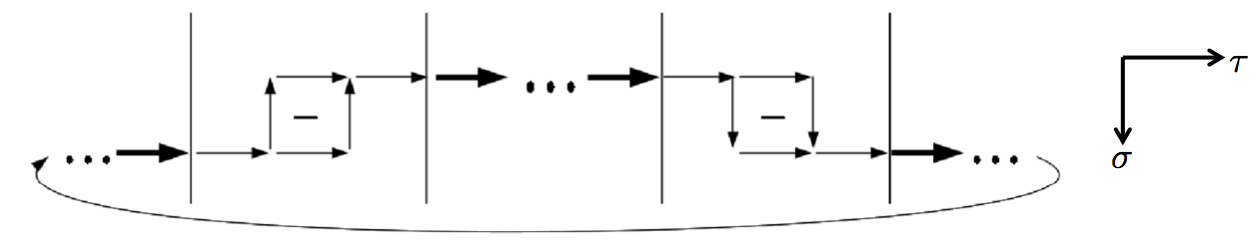}
\caption{Lattice discretization of the colour-electric correlator Eq.~(\ref{eq:hqlat})}
\label{fig:hqlat}
\end{figure}

Due to the difficulties encountered in disentangling the dissociation and transport regions when reconstructing spectral functions from lattice correlators, it is advantageous to find lattice accessible correlators that couple exclusively to one of the two. For heavy quark diffusion such an exclusive lattice correlator was derived in \cite{CaronHuot:2009uh,CasalderreySolana:2006rq}. It describes a single heavy quark propagating the thermal medium:
\begin{equation}
 G(\tau) =
 \frac{
  \Big\langle
   {\rm Re Tr} \Big[
      U(\frac{1}{T};\tau) \, gE(\tau,\vec{0}) \, U(\tau;0) \, gE(0)
   \Big] 
  \Big\rangle
 }{3
 \Big\langle
    {\rm Re Tr}[U(\frac{1}{T};0)] 
 \Big\rangle~~.
 }
 \label{eq:hqlat}
\end{equation}
Its spectral function is related to heavy quark momentum diffusion $\kappa$ and the diffusion constant $D$ via
\begin{equation}
 \kappa=\lim_{\omega\rightarrow 0}\frac{2T\rho_{\rm E}(\omega)}{\omega}~,~~~ D=\frac{2T^2}{\kappa}~~.
\end{equation}
This HQET,  colour-electric correlator is a purely gluonic quantity and powerful methods exist to boost its signal \cite{Luscher:2001up,Meyer:2007ic,Parisi:1983hm,DeForcrand:1985dr}.
In addition, there is freedom to choose an optimal discretization for the electric field insertions \cite{Meyer:2010tt,Banerjee:2011ra}. The pictoral representation of one possible choice \cite{Francis:2011gc,Francis:2013cva,Kaczmarek:2014jga} is given in Fig.~\ref{fig:hqlat}.

\begin{figure}[t!]
\centering
\includegraphics[width=0.4\textwidth]{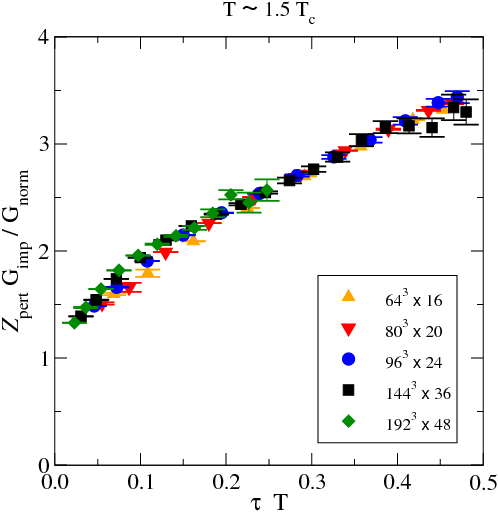}
\includegraphics[width=0.42\textwidth]{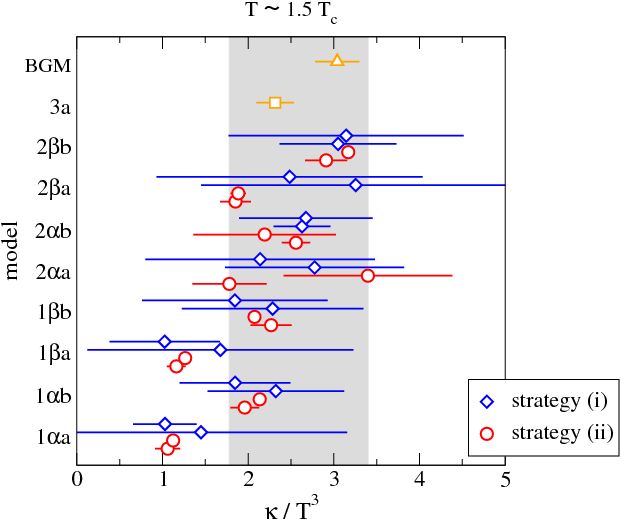}
\caption{Top: Lattice results for the perturbatively renormalized, tree-level improved colour-electric correlator for five different lattice spacings \cite{Francis:2015daa}. Bottom: The heavy quark momentum diffusion coefficients estimated after spectral function reconstruction using multiple ans\"atze and the Backus-Gilbert method (BGM). The grey band denotes the final result.}
\label{fig:hqdiff}
\end{figure}

Using this discretization a reconstructed result in the continuum limit of quenched QCD has now become available \cite{Francis:2015daa}. The authors carried out a series of calculations with lattice spacings varying from $a=0.03$fm to $0.010$fm on large, isotropic, quenched ensembles, utilizing the multi-level algorithm \cite{Luscher:2001up} and semi-analytic link integration \cite{DeForcrand:1985dr} to achieve good signal for lattice time extents of up to $N_\tau=48$. Renormalization was carried out via a 1-loop perturbative computation and the data was tree-level improved using $G_{\rm cont}^{\rm LO}(\overline{\tau T}) = G_{\rm lat}^{\rm LO}(\tau T)$. The central results are shown in Fig.~\ref{fig:hqdiff}(top), whereby the exponential decay of the correlator was cancelled off using the leading order result $G_{\rm norm}(\tau)$. The continuum limit was achieved using a combined resampling and b-spline interpolation of the data at fixed $\tau T$ in $1/N_\tau^2$. The spectral function reconstruction was carried out using several interpolating ans\"atze between the known IR and UV asymptotics:
	\begin{equation}
	\phi_{\rm IR}(\omega)=\frac{\kappa\omega}{2T} ~,~
	\phi_{\rm UV}(\omega)= \frac{g^2({\rm max(\omega,\pi T)})C\omega^3 }{6\pi}
	\end{equation}
In addition the spectral functions were reconstructed using the Backus-Gilbert method \cite{Backus:1968bg} as cross-check. The resulting transport coefficients from these different reconstructions are collected in Fig.~\ref{fig:hqdiff}(bottom). Based on the central values from this figure the heavy quark momentum diffusion is estimated to be $\kappa/T^3 \sim 1.8...3.4$.

This result can be turned into an estimate of the kinetic equilibration time \cite{Francis:2015daa} via
\begin{equation}
 \tau_{\rm kin} = \frac{1}{\eta_D}=(1.8...3.4)\Big( \frac{T_c}{T}\Big)^2\,\Big( \frac{M}{1.5 \rm GeV} \Big)\, {\rm fm}/c~.
\end{equation}
As a consequence, close to $T_c$ charm quark kinetic equilibration seems to be 
almost as fast as that of light partons, where a time scale
$\sim 1$~fm/c is often quoted.

\section{Summary}

In summary, we have recently seen the first calculations of the complex heavy quark potential. Nevertheless additional work is needed to pin down the imaginary part in dynamical QCD. Aside of new quenched results, dynamical calculations of charmonia at finite temperature are underway. There is still some uncertainty whether the $J/\Psi$ dissociates before $1.5T_c$ or survives to higher temperatures. The lattice calculations using the finest and largest ensembles observe the lower dissociation temperatures. This discrepancy might be due to lattice artifacts or the reconstruction method. Future studies attempting a continuum limit with more varied reconstruction methods will hopefully shed light on the origin of this discrepancy.
Several NRQCD studies of bottomonia agree that they do not observe any dissociation of the $\Upsilon$ below $1.6T_c$. Additionally quenched studies with cut-offs large enough to accommodate a fully relativistic approach to bottomonia are underway. These results will also enable an independent determination of bottom quark diffusion. Finally,
the heavy quark momentum diffusion coefficient has been calculated in the continuum limit of quenched QCD at $1.5T_c$. The values found are $\kappa/T^3 \sim 1.8...3.4$ and a simple estimate of the kinetic equlibration time $\tau_{\rm kin}$ points to heavy quarks equilibrating almost as fast as light ones.


 \section*{Acknowledgments}
I wish to thank the organizers of ”Hard Probes” for giving me the opportunity to present this review and for hosting such a stimulating conference. Additionally I am grateful to G. Aarts, M. Kitazawa, H. Ohno and O. Kaczmarek for kindly sharing their results. Finally, I thank B. B. Brandt for looking over this draft.


\nocite{*}
\bibliographystyle{elsarticle-num}
\bibliography{Francis_A.bib}







\end{document}